\documentclass[12pt,draftcls,onecolumn]{IEEEtran}   

\usepackage{times}
\usepackage{epsfig}
\usepackage{times}
\usepackage{color}
\usepackage{amsfonts}
\usepackage{amssymb}
\usepackage{amsbsy} 
\usepackage{amsmath}
\usepackage[ansinew]{inputenc}
\usepackage{pstricks}
\usepackage{lscape}
\usepackage{cite}
\usepackage{setspace}
\usepackage{rotating}
\usepackage{multirow}
\usepackage[hidelinks]{hyperref}

\renewcommand{\green}[1]{\textcolor[rgb]{0,0.5,0}{#1}}           

\newcommand{\x}{{\mathbf x}}
\newcommand{\f}{{\mathbf f}}

\newcommand{\y}{{\mathbf y}}
\newcommand{\Y}{{\mathbf Y}}
\newcommand{\X}{{\mathbf X}}

\newcommand{\IG}{\includegraphics}

\usepackage{hyperref}
\hypersetup{
    bookmarks=true,         
    unicode=false,          
    pdftoolbar=true,        
    pdfmenubar=true,        
    pdffitwindow=false,     
    pdfstartview={FitH},    
    pdftitle={GALGA},    
    pdfauthor={Jorge Vicent},     
    pdfsubject={GALGA},   
    pdfcreator={Jorge Vicent},   
    pdfproducer={Jorge Vicent}, 
    pdfkeywords={keyword1} {key2} {key3}, 
    pdfnewwindow=true,      
    colorlinks=true,       
    linkcolor=blue,          
    citecolor=blue,        
    filecolor=blue,      
    urlcolor=blue           
}

\begin{document}

\title{Gradient-based Automatic Look-Up Table Generator for Atmospheric Radiative Transfer Models}

\author{
Jorge Vicent, Luis Alonso, Luca Martino, Neus Sabater, Jochem Verrelst, Gustau Camps-Valls,~\IEEEmembership{Fellow,~IEEE} and Jos\'e Moreno,~\IEEEmembership{Senior Member,~IEEE}

\thanks{{\bf Manuscript published in IEEE Transactions on Geoscience and Remote Sensing, vol. 57, no. 2, pp. 1040-1048, Feb. 2019, doi: 10.1109/TGRS.2018.2864517.} This work was carried out in the frame of ESA's project {\it FLEX L2 End-to-End Simulator Development and Mission Performance Assessment} ESA Contract No. 4000119707/17/NL/MP. Jochem Verrelst was supported by the European Research Council (ERC) under the ERC-2017-STG SENTIFLEX project (grant agreement 755617). Gustau Camps-Valls and Luca Martino were supported by the ERC under the ERC-CoG-2014 SEDAL project (grant agreement 647423).}
\thanks{J. Vicent, L. Alonso, L. Martino, N. Sabater, J. Verrelst, G. Camps-Valls and J. Moreno are with Image Processing Laboratory, University of Valencia, 46980 Paterna (Valencia), Spain (e-mail: [\href{mailto:jorge.vicent@uv.es}{jorge.vicent}; \href{mailto:luis.alonso@uv.es}{luis.alonso}; \href{mailto:luca.martino@uv.es}{luca.martino}; \href{mailto:m.neus.sabater@uv.es}{m.neus.sabater}; \href{mailto:jochem.verrelst@uv.es}{jochem.verrelst}; \href{mailto:gustau.camps@uv.es}{gustau.camps}; \href{mailto:jose.moreno@uv.es}{jose.moreno}]@uv.es).}
}

\maketitle

\begin{abstract}
Atmospheric correction of Earth Observation data is one of the most critical steps in the data processing chain of a satellite mission for successful remote sensing applications. 
Atmospheric Radiative Transfer Models (RTM) inversion methods are typically preferred due to their high accuracy. However, the execution of RTMs on a pixel-per-pixel basis is impractical due to their high computation time, thus large multi-dimensional look-up tables (LUTs) are precomputed for their later interpolation. 
To further reduce the RTM computation burden and the error in LUT interpolation, we have developed a method to automatically select the minimum and optimal set of nodes to be included in a LUT. 
We present the gradient-based automatic LUT generator algorithm (GALGA) which relies on the notion of an acquisition function that incorporates (a) the Jacobian evaluation of an RTM, and (b) information about the multivariate distribution of the current nodes.
We illustrate the capabilities of GALGA in the automatic construction and optimization of MODerate resolution atmospheric TRANsmission (MODTRAN) LUTs for several input dimensions. 
Our results indicate that, when compared to a pseudo-random homogeneous distribution of the LUT nodes, GALGA reduces (1) the LUT size by $\sim$75\% and (2) the maximum interpolation relative errors by 0.5\%. 
It is concluded that automatic LUT design might benefit from the methodology proposed in GALGA to reduce computation time and interpolation errors. 
\end{abstract}


\section{Introduction}

Atmospheric correction of Earth Observation data aims to derive surface properties (e.g., reflectance) through the inversion of the atmospheric radiative transfer equations. It is perhaps one of the most critical steps in the data processing chain of a satellite mission for successful remote sensing applications~\cite{Schaepman2009}. Though {\it empirical} atmospheric correction methods~\cite{ChavezJr19961025} typically have a low computation burden, {\it physically-based} methods~\cite{Cooley2002,Richter2002,Guanter2007709,North2008} are often preferred as their accuracy is generally higher~\cite{Bernstein2012,Kokhanovsky2007}. These {\it physically-based} methods rely on the inversion through a Radiative Transfer Model (RTM)~\cite{Vermote1997675,Berk2006}, which are however computationally expensive and very often impractical for their execution on a pixel-per-pixel basis~\cite{CampsValls11mc}. To overcome this limitation, large multi-dimensional look-up tables (LUTs) are precomputed for their later interpolation~\cite{Guanter2009}. However, little information is available in the scientific literature about the criteria that should be adopted to design these LUTs, and about the errors derived of their interpolation. In addition, the computation of these LUTs is still time consuming, requiring techniques of parallelization and execution in computer grids~\cite{Brazile2008,Huang2016}.
 
In order to further reduce the RTM computation time, a possible strategy is to select the minimum and optimal set of points (nodes, anchors) to be included in a LUT that reduce the error in its interpolation. 
This problem is known as {\em experimental optimal design}~\cite{Chaloner95,DANI15} of interpolators of arbitrary functions $f$, and it aims at reducing the number of direct evaluations of $f$ (RTM runs in the context of LUT design). A possible approach is to construct an approximation of $f$ starting with a set of initial points. This approximation is then sequentially improved incorporating new points given a suitable selection rule until a certain stop condition is satisfied.  
Another interesting alternative approach is based on {\em adaptive gridding}, which aims to construct a partitioning of the input variable space, $\mathcal{X}$, into cells of equal size, where the cell edges have different lengths depending on their spatial direction~\cite{Busby09}. In order to find such lengths, the {\em adaptive gridding} method uses a Gaussian Process (GP) model with an automatic relevant determination kernel~\cite{Rasmussen05,CampsValls16grsm}. A clear problem of such approach is that the number of hyper-parameters to be estimated increases as the input dimension grows. 
The topic of {\em experimental optimal design} has received attention from (apparently unrelated) research areas such as optimal nonuniform sampling, quantization and interpolation of continuous signals~\cite{Marvasti01}, Bayesian Optimization (BO)~\cite{Gutmann15,Mockus89}, and active learning~\cite{Verrelst16gpal}.

The main objective of this paper is, therefore, to present a simpler method for the automatic generation of RTM-based LUTs. As a proof of concept, the proposed methodology is applied to the widely MODerate resolution atmospheric TRANsmission (MODTRAN) RTM for the generation of atmospheric LUTs. The ultimate goal is thus to reduce errors in the RTM LUT interpolation and thus in the atmospheric correction of Earth Observation data. The proposed method is sequential and automatically builds the LUT based on the notion of the {\it acquisition function}, similarly to the BO approach~\cite{Gutmann15,Mockus89}. Such acquisition function acts as a sort of oracle that tells us about the regions of the space more interesting or informative to sample. Essentially, starting from a set of initial points, the LUT is therefore built automatically with the addition of new nodes maximizing the acquisition function at each iteration. Unlike in BO, our goal is not the optimization of the unknown underlying function $f$ but its accurate {\it approximation} $\widehat f$ through minimization of its interpolation error $\delta$. Thus, the {\em experimental optimal design} problem is converted into a {\em sequential optimization} problem of the acquisition function, regardless of the dimensionality of the input space.

The remainder of the present work is structured as follows. Section \ref{sec:method} details the implemented gradient-based automatic LUT generator algorithm. Section \ref{sec:simsetup} describes the experimental simulation set-up including the methodology to evaluate the performance of the proposed algorithm. Section \ref{sec:results} shows the functioning of the algorithm and its performance for LUTs of different dimensionality. Finally, in Section~\ref{sec:conclusions}, we conclude our work with a discussion of the results in the context of atmospheric correction for Earth Observation applications, and an outlook of future research lines.

\section{Gradient-based automatic LUT generator}
\label{sec:method}

This section describes the developed gradient-based automatic LUT generator algorithm (GALGA). We start in Section \ref{ssec:overview} by giving a schematic overview of the proposed algorithm and the employed notation. We then detail in Sections \ref{ssec:interpolation}, \ref{ssec:stopcond} and \ref{ssec:acquisition} the specificities of the algorithm through the implemented {\it interpolation} and the concepts of the {\it acquisition function} and the {\it stop condition}.

\subsection{Method overview}
\label{ssec:overview}

\begin{figure}[h!]
	\centering
	\IG[width=\linewidth,trim=1.2cm 1.2cm 1.2cm 0.9cm,clip=true]{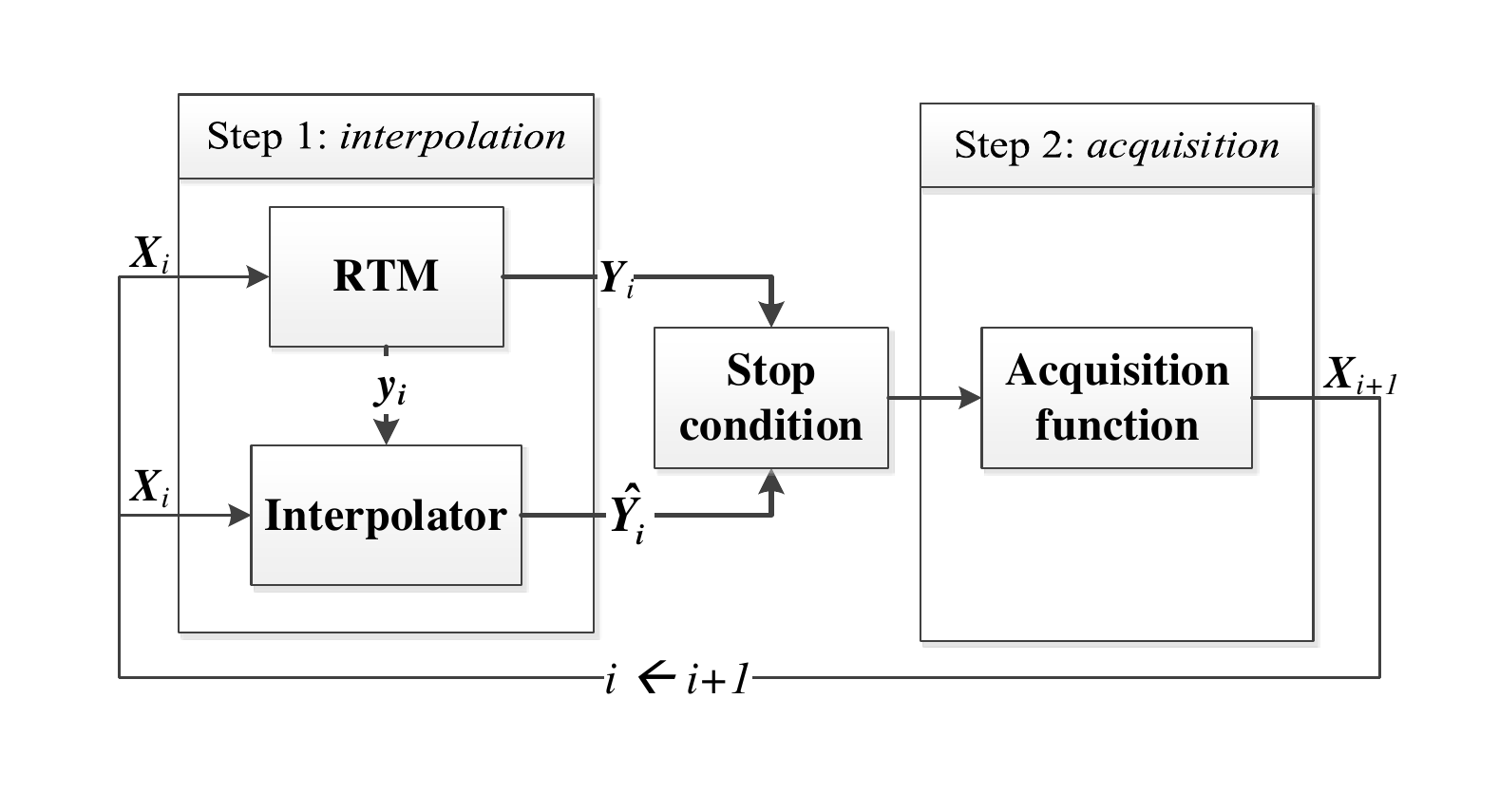}
	\caption{Schematic representation of GALGA's processing chain.}
	\label{fig:scheme}
\end{figure}

The basic component of GALGA is the acquisition function based on geometric and density terms, and was originally introduced in~\cite{luca2017a,CampsValls17scia}. See Fig. \ref{fig:scheme} for an illustrative processing scheme of the method. Notationally, let us consider a $D$-dimensional input space $\mathcal{X}$, i.e., $\x \in\mathcal{X}\subset \mathbb{R}^D$ in which a costly $K$-dimensional object function $\f(\x;\lambda) = [f(\x;\lambda_1),\ldots,f(\x;\lambda_K)]: \mathcal{X}\mapsto \mathbb{R}^K$ is evaluated. In the context of this paper, $\mathcal{X}$ comprises the input space of atmospheric and geometric variables (e.g., Aerosol Optical Thickness (AOT), Visual Zenith Angle (VZA)) that control the behavior of the function $\f(\x;\lambda)$, i.e., an atmospheric RTM. Here, $\lambda$ represents the wavelengths in the $K$-dimensional output space. For sake of simplicity, this wavelength dependency is omitted in the formulation in this paper, $\f(\x;\lambda)\equiv\f(\x)$. 
Given a set of input variables in the matrix $\X_i=[\x_1,\ldots,\x_{m_i}]$ of dimension $D\times m_i$, we have a matrix of $K$-dimensional outputs $\Y_i=[\y_1,\ldots,\y_{m_i}]$, being $\y_j=\f(\x_j)$ for $j\in$[1, $m_i$].
At each iteration $i\in \mathbb{N}^+$, GALGA first performs an {\em interpolation}, $\widehat{\y}_i\equiv\widehat{\f}_i(\x|\X_i,\Y_i)$, of the function $\f(\x)$. Second, the algorithm follows with an {\em acquisition} step that creates/updates the acquisition function, $A_i(\x)$, and increases the number of LUT nodes from [$\X_i$;$\Y_i$] to $\X_{i+1} =[\x_1,\ldots,\x_{m_{i+1}}]$ and $\Y_{i+1}=[\y_1,\ldots,\y_{m_{i+1}}]$. This two-steps procedure is repeated until a suitable stopping condition is met based on the difference between $\f(\x)$ and $\widehat{\f}_i(\x)$.

The algorithm starts ($i$=0) by choosing $N_0=5\cdot2^D$ pseudo-random nodes based on a Latin Hypercube Sampling~\cite{McKay1979} of the input variable space. This initial set of LUT nodes is complemented with the addition of all the 2$^D$ vertex of the input variable space (where the input variables get the minimum/maximum values). With this set of $m_0=N_0+2^D$ nodes, we ensure to have an initial homogeneous and bounded distribution of the input variable space so that no extrapolations are performed.

\subsection{Interpolation method}
\label{ssec:interpolation}

\begin{figure}[h!]
	\centering
	\IG[width=\linewidth]{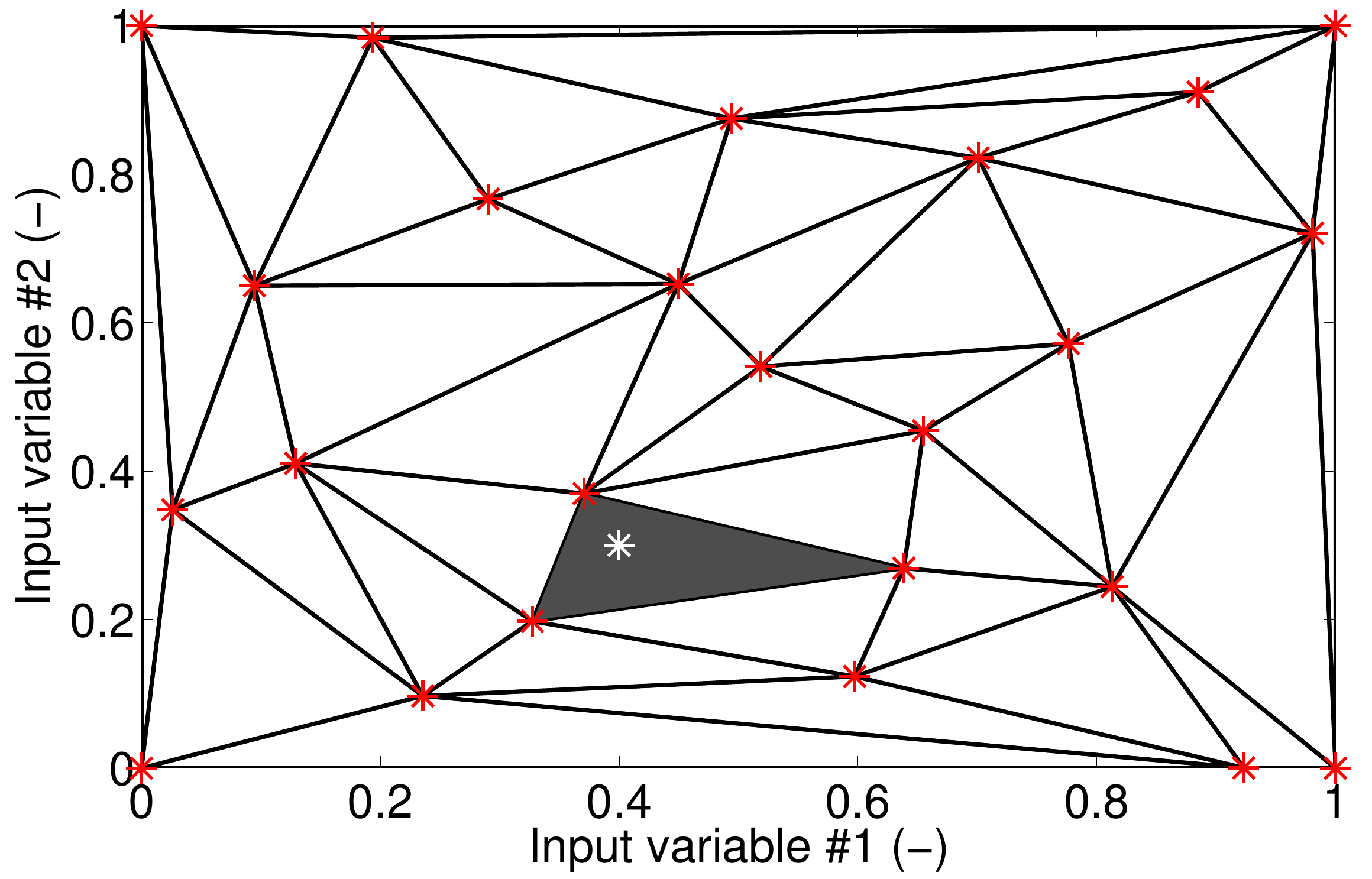}
	\caption{Schematic representation of a $2$-dimensional interpolation of a query point $\x_q$ (white $\ast$) after Delaunay triangulation (solid lines) of the scattered LUT nodes $X_i$ ({$\ast$}).}
	\label{fig:delaunay}
\end{figure}

GALGA relies on the use of an interpolation method $\widehat \f(\x)$ in order to provide an approximation of the underlying function $\f(\x)$ within $\mathcal{X}$.
{In our previous work~\cite{luca2017a,CampsValls17scia}, we considered a GP interpolator~\cite{Rasmussen05}, widely used in various remote sensing applications~\cite{CampsValls16grsm}. Interpolation in GPs is trivially implemented by considering that there is no noise in the observed signal, and hence only the kernel hyperparameters need to be learned. However, the use of GP for multi-output functions (i.e., $K>1$) is not straighforward, which most of the times requires conducting first a dimensionality reduction~\cite{Verrelst2015,Verrelst2016} followed by individual GPs for each projection. Not only the model complexity increases, but also the risk of falling in local minima because of the problems of learning hyperparameters in multiple GP models. 
In GALGA, we instead implemented a multi-dimensional linear interpolation method, commonly applied in physically-based atmospheric correction methods~\cite{Cooley2002,Richter2002,Guanter2009}.}
The implementation of the linear interpolation is based on MathWorks' MATLAB function \texttt{griddatan}, which relies on the Quickhull algorithm~\cite{Barber1996469} for triangulations in multi-dimensional input spaces. 
For the scattered input data in $\X_i$, the linear interpolation method is reduced to find the corresponding Delaunay's simplex~\cite{Delaunay1934} (e.g., a triangle when $D=2$) that encloses a query $D$-dimensional point $\x_q$ (see Fig. \ref{fig:delaunay}):
\begin{equation} \label{eqn:lininterp}
\widehat{\f}_i(\x_q) = \sum_{j=1}^{D+1}\omega_j\f(\x_j),
\end{equation}
where $\omega_j$ are the (scalar) barycentric coordinates of $\x_q$ with respect to the $D$-dimensional simplex (with $D+1$ vertices)~\cite{coxeter}.

Since $\f(\x)$ is a $K$-dimensional function, the result of the interpolation is also $K$-dimensional. The Delaunay triangulation, in turn, provides partitions of the input space in simplices. The use of these simplices will help us to define the {\it acquisition function} (see Section \ref{ssec:acquisition}).

\subsection{The stop condition}
\label{ssec:stopcond}

\begin{figure}[h!]
	\centering
	\IG[width=\linewidth]{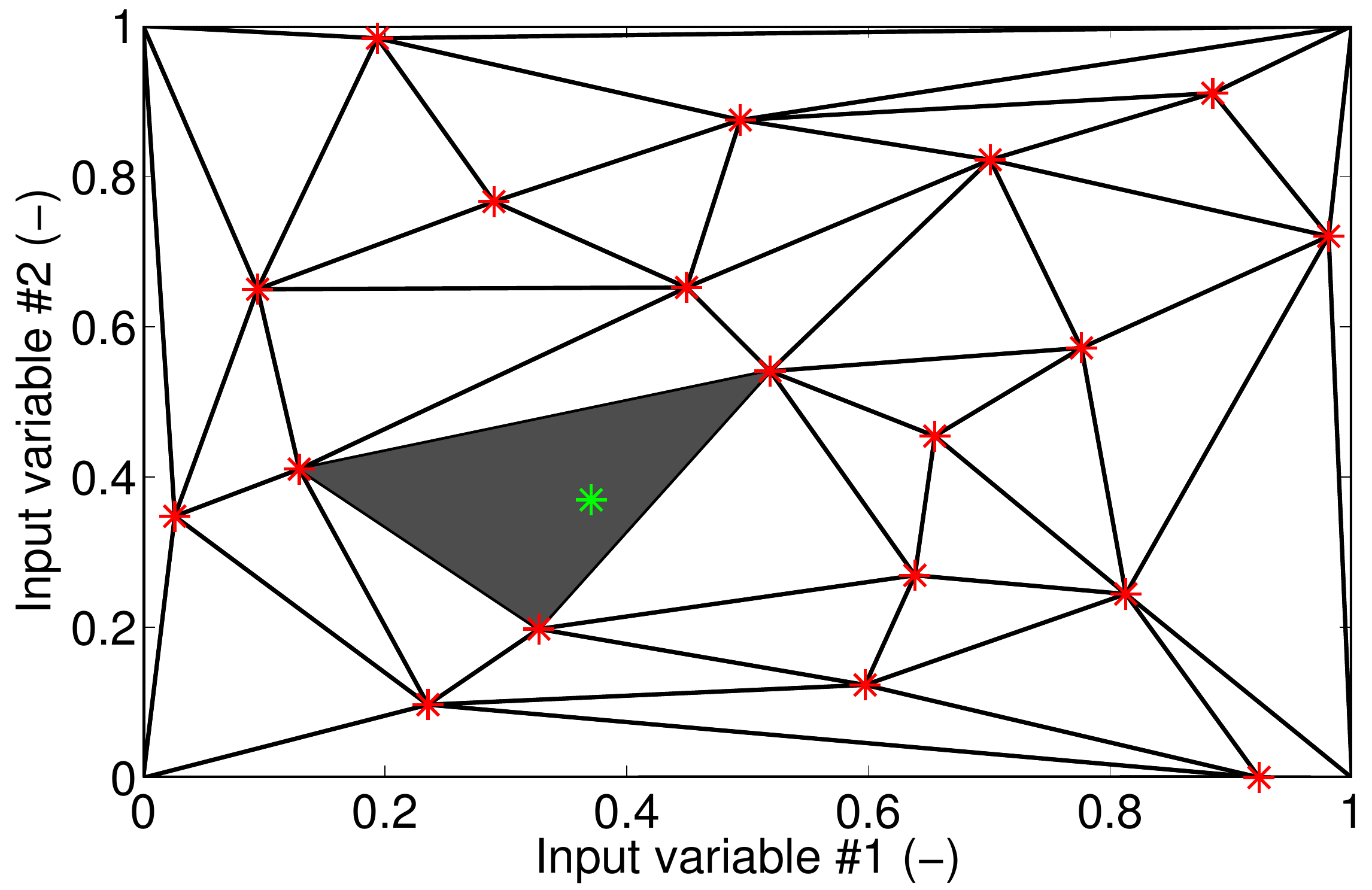}
	\caption{Schematic process for the calculation of $\delta_i$ in equation \eqref{eqn:errormetric} within the $\bar\X_i$ subset (colored $\ast$). Notice how the ``leave-one-out'' cross-validation technique modifies the Delaunay simplices with respect the complete $\bar\X_i$ subset in Fig. \ref{fig:delaunay}.}
	\label{fig:stopcond}
\end{figure}

The purpose of the stop condition is to end the iterative process of the algorithm when a suitable condition in the LUT data is met. In the proposed algorithm, the stop condition is based on the evaluation of the interpolation error through the error metric $\delta_i(\x)$
\begin{equation} \label{eqn:errormetric}
\delta_i(\bar\X_i) = \max_\lambda\left(100\cdot\left|\frac{\widehat{\f}_i(\bar\X_i) - \f(\bar\X_i)}{\f(\bar\X_i)}\right|\right),
\end{equation}
where $\bar\X_i$ is a subset of $\X_i$ that comprises all the LUT nodes at the $i$-th iteration with the exception of the 2$^D$ vertex of the input variable space. The error metric, therefore, evaluates the interpolation relative error over each node in the subset $\bar\X_i$ by using the leave-one-out cross-validation technique (see the green $\ast$ in Fig. \ref{fig:stopcond})~\cite{Hastie09}. Among all the spectral channels ($\lambda$), this error metric takes the most critical spectral channel ($\max_\lambda$). The iterative process finishes when the 95\% percentile of $\delta_i(\bar\X_i)$ is below an error threshold, $\varepsilon_t$.

By taking the spectral channel under which the interpolation relative error is maximum, the stop condition ensures that all the spectral channels will have an interpolation error lower than $\varepsilon_t$. In this way, GALGA will be valid for (and independent of) all remote sensing applications. With respect to the error threshold, this can be user-defined according to some pre-calculated condition as e.g., a factor 10 over the instrument absolute radiometric accuracy.  

It should be noted that the leave-one-out cross-validation technique does not provide the ``true'' error of the interpolation over all the input space $\mathcal{X}$ but an approximation. Since the cross-validation technique leaves some LUT nodes out of the LUT, it is expected that the calculated interpolation relative error in equation \eqref{eqn:errormetric} will be higher than the ``true'' error. However, as the LUT nodes are also used to determine the interpolation error, using this cross-validation technique allows us to avoid generating an external (i.e., not included in the final LUT) validation dataset.

\subsection{The acquisition function}
\label{ssec:acquisition}
The acquisition function, $A_i(\x)$, is the core of the proposed algorithm since it allows determining the new LUT nodes to be added at each iteration. This function incorportates (a) {\it geometric} information of the unknown function $\f$ through the evaluation of its gradient, and (b) {\it density} information about the distribution of the current nodes. Indeed, areas of high variability of $\f(\x)$ require the addition of more LUT nodes as well as areas with a small concentration of nodes require the introduction of new inputs. Accordingly, we define the acquisition function conceptually in equation \eqref{eqn:acqfunc} as the product of two functions: a {\it geometric term} $G_i(\x)$ and a {\it density term} $D_i(\x)$:
\begin{equation} \label{eqn:acqfunc}
A_i(\x) = G_i(\x)^{\beta_i}D_i(\x)^{1-\beta_i},
\end{equation}
where $\beta_i$ is a discrete function that alternates the acquisition function between the geomety and density terms every $T=3$ iterations:\begin{equation} \label{eqn:beta}
\beta_i = \beta_{i+T} =
\begin{cases}
	1	& \quad \text{if } i\leq T-1 \\
	0 	& \quad \text{if } i=T.\\
\end{cases}
\end{equation}
The {\it geometric term} $G_i(\x)$ is based on the calculation of the gradient of the underlying function $\f$. However, since $\f$ is unknown in all the input variable space $\mathcal{X}$, the gradient can only be approximated and calculated at the current LUT nodes $\X_i$. Therefore, $G_i(\x)$ is calculated according to the following steps, as shown in Fig. \ref{fig:geometryfunc}:
\begin{figure}[h!]
	\centering
	\IG[width=\linewidth]{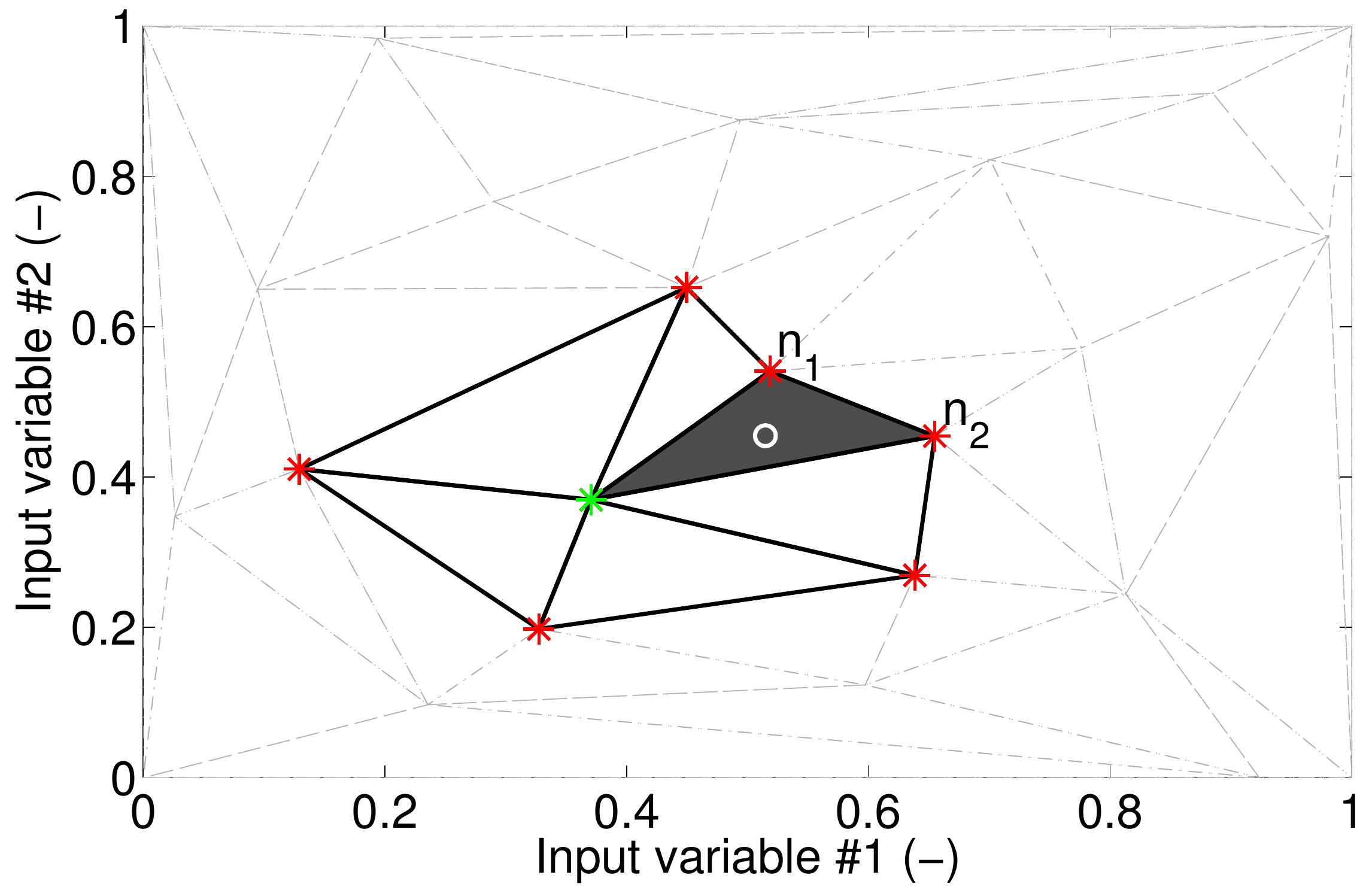}
	\caption{Schematic representation for the calculation of $G_i(\x)$. Gradients are calculated between the nodes $\x_k$ ({$\ast$}) and the selected LUT node $\x_j$ (green $\ast$). A new LUT node (white $\circ$) is added at the barycenter of the Delaunay simplex with highest average gradient (shaded in dark grey).}
	\label{fig:geometryfunc}
\end{figure}

\begin{enumerate}
	\item Among the LUT nodes in $\X_i=[\x_1,\ldots,\x_{m_i}]$, we select only those $m_{g,i}$ nodes whose interpolation error $\delta_i(\x_j)$ (see Eq. \eqref{eqn:errormetric}) is higher than the error threshold $\varepsilon_t$. By choosing this subset, the new LUT nodes will only be added in areas with high interpolation error.
	\item The gradient, $\nabla_k \f_i(\x_j)$, is calculated according to equation \eqref{eqn:gradient} between the current node $\x_j$ (\green{$\ast$} in Fig.  \ref{fig:geometryfunc}) and all the $N_k$ remaining nodes ($\x_k$ with $k\in[1,N_k]$) of the Delaunay simplices for which $\x_j$ is a vertex ({$\ast$} in Fig. \ref{fig:geometryfunc}):
		\begin{equation} \label{eqn:gradient}
			\nabla_k \f_i(\x_j)_{\lambda_{max}} = \left|\y_j - \y_k\right|_{\lambda_{max}} ,
		\end{equation}
The subindex $\lambda_{max}$ indicates that, out of the $K$-dimensional output values in $\y$, only the most critical spectral channel (see Section \ref{ssec:stopcond}) is used to calculate the gradient. 
	\item For each Delaunay simplex ($l$), we calculate the root-mean-square of the corresponding $D$ gradients in the previous step according to equation \eqref{eqn:gradient2}:
		\begin{equation} \label{eqn:gradient2}
			g_l = \sqrt{\frac{1}{D}\sum_{n=n_1}^{n_D}\left(\nabla_n \f_i(\x_j)_{\lambda_{max}}\right)^2}, 
		\end{equation}
		where the index $n$ (from $n_1$ to $n_D$) identifies the $D$ nodes, among $\x_l$, that conform a Delaunay simplex together with $\x_j$ (see $n_1$ and $n_2$ tagged nodes in Fig. \ref{fig:geometryfunc}).
	\item The gradient term finally adds a new LUT node at the barycenter of the Delaunay simplex with higher value of $g_l$.
\end{enumerate}
Following the previous steps, $G_i(\x)$ will place a new node in the vicinity of each current LUT node in $\X_i$ with an interpolation error higher than $\varepsilon_t$ in the direction of the highest gradient. Therefore, the LUT size will increase from $m_i$ nodes to $m_{i+1}=m_i + m_{g,i}$ nodes.

Since the {\it gradient term} is based on the existing LUT nodes ($\X_i$), the computed interpolation errors and gradients might not be representative in empty areas of the input variable space, particularly in those with low density of nodes. Thus, the acquisition function includes a {\it density term}, $D_i(\x)$, which aims at proofing these lower sampled areas every $T$ iterations (see equations \eqref{eqn:acqfunc} and \eqref{eqn:beta}). The {\it density term} identifies these poorly sampled areas by calculating the volume of each Delaunay simplex according to the equation \eqref{eqn:volsimplex}~\cite{stein1966}:
\begin{equation} \label{eqn:volsimplex}
	V = \frac{1}{D!}\det\left(\x_{n_2}-\x_1, \ldots,\x_{n_{D+1}}-\x_{n_1} \right), 
\end{equation}
where the indices $n_1$ to $n_{D+1}$ identify the $D+1$ nodes that conform each $D$-dimensional Delaunay simplex. 
The {\it density term} will then place a new LUT node in the barycenter of the $m_{d,i}=5\cdot2^D$ simplices with higher volume. Therefore, the LUT size will increase from $m_i$ nodes to $m_{i+1}=m_i + m_{d,i}$ nodes.

\section{Experimental set-up and analysis}
\label{sec:simsetup}

In order to analyze the functioning and performance of the proposed algorithm, we run three simulation test cases for the optimization of MODTRAN5-based LUTs. MODTRAN5 is one of the most widely used atmospheric RTM for atmospheric correction applications due to its accurate simulation of the coupled absorption and scattering effects~\cite{Berk1998367,Berk2006}. Following the notation in Section \ref{sec:method}, the underlying function $\f$ consists of Top-Of-Atmosphere (TOA) radiance spectra, calculated at a Solar Zenith Angle (SZA), $\theta_{il}$, and for a Lambertian surface reflectance\footnote{We consider the conifer trees surface reflectance from ASTER spectral library~\cite{Baldridge2009711}}, $\rho$, according to equation \eqref{eqn:lambert}:
\begin{equation} \label{eqn:lambert}
	L = L_0 + \frac{(T_{dir}+T_{dif})(E_{dir}\cos\theta_{il}+E_{dif})\rho}{\pi(1-S\rho)},
\end{equation}
where $L_0$ is the path radiance, $T_{dir/dif}$ are the target-to-sensor direct/diffuse transmittances, $E_{dir/dif}$ are the direct/diffuse at-surface solar irradiances and $S$ is the spherical albedo. These terms are often called {\it atmospheric transfer functions} and are obtained using the MODTRAN5 interrogation technique developed in~\cite{Guanter2009}. Unless otherwise specified, all simulations are carried out for a nadir-viewing satellite sensor (VZA=0.5 deg), target at 0 km altitude, rural aerosols and mid-latitude summer atmosphere.
 
The three simulation test cases consist of LUTs of increasing dimensionality of the input space i.e., $D=[2; 4; 6]$, in the wavelength range 400-550 nm at 15 cm$^{-1}$ spectral sampling ($\approx$0.4 nm). The input variables (see Tab. \ref{tab:scenario1}) range typical variability in the AOT, the \r{A}ngstr\"om exponent ($\alpha$), the Henyey-Greenstein asymmetry parameter ($g_{HG}$) and the single scattering albedo (SSA)~\cite{Holben1998,Hess1998831,Dubovik2002590}.

\begin{table}[h]
\small
\centering
\caption{Input variables and spectral configuration for the visualization test scenario. For cases \#2 and \#3, SZA takes a constant value of 55 deg.}
\begin{tabular}{c l c}
\textbf{Case}					& \textbf{Input variables (range)}	 	& \textbf{Error threshold, $\mathbf \varepsilon_t$ (\%)}	\\  \hline
\multirow{ 2}{*}{\textbf{\#1}}		& AOT (0.05-0.4)					& 0.2 					 					\\ 
							& SZA (20-70 deg)					& 											\\ \hline
\multirow{ 2}{*}{\textbf{\#2}}		& As in Case \#1 plus...				& 1											\\ 
							& $\alpha$ (1-2)					& 											\\ 
							& $g_{HG}$ (0.60-0.99)				& 											\\ \hline
\multirow{ 2}{*}{\textbf{\#3}}		& As in Case \#2 plus...				& 2											\\ 
							& SSA (0.85-0.99)					& 											\\ 
							& VZA (0.5-20 deg)					& 											\\ 
\hline
\end{tabular}
\label{tab:scenario1}
\end{table}

We start the analysis of the data by {\bf visualizing the functioning} of the algorithm in terms of: (1) the evaluation of the stop condition through cross-validation error, and (2) the distribution of new nodes according to $G_i$ and $D_i$. To do so, we exploit the $2$-dimensional data in Case \#1, showing the {\it cross-validation} and the {\it true} error maps. These two maps are shown at two consecutive iterations, which correspond to the actuation of each term ({\it geometry} and {\it density}) of the acquisition function. 
On the one hand, the {\it cross-validation} error maps are based on the $\delta_i$ (see equation \eqref{eqn:errormetric}) calculated through the ``leave-one-out'' cross-validation of each subset $\bar\X_i$ as introduced in Section \ref{ssec:stopcond}. To create a bi-dimensional map, the scattered values of  $\delta_i(\bar\X_i)$ are linearly interpolated over a grid of 100$\times$100 linearly-spaced values of the input variables. 
Since this cross-validation method reduces locally the LUT nodes density (thus the name ``leave-one-out''), the resulting error maps should not be understood as an estimation of the underlying LUT interpolation errors. Instead, the purpose of the {\it cross-validation} error maps is to illustrate the distribution and magnitude of the cross-validation errors, which are the ones used to determine the distribution of new LUT nodes. Overlapped with these error maps, the current LUT nodes $\X_i$ and their Delaunay triangulation are shown together with the nodes added at the iteration $i+1$.
On the other hand, the {\it true} error maps correspond to the $\delta_i$ calculated over a grid of 100$\times$100 linearly-spaced values of the input variables where TOA radiance spectra is pre-calculated. Namely, this thin grid represents the true value of $\f(\x)$.\\

We continue the analysis of the data by {\bf assessing the performance} of the proposed algorithm in the test cases \#1, \#2 and \#3. For each test case, we calculate (1) the 95\% percentiles ($P_{95}$) of $\delta_i$ obtained from the cross-validation subset $\bar\X_i$ and (2) the values of $P_{95}$, $P_{97.5}$ and $P_{100}$ (i.e., maximum error) from the $\delta_i$ calculated with a reference ({\it ground truth}) LUT. 
These {\it ground truth} LUTs consist on nearly 13'000, 32'000 and 72'000 nodes, respectively for cases \#1, \#2 and \#3, homogeneously distributed in the input variable space according to a Latin Hypercube Sampling.
Since the initial node distribution in our algorithm is pseudo-random, we calculate the mean and standard deviation of $P_{95}$ in the cross-validation subset after 10 independent runs. The performance of the proposed algorithm is shown by plotting these statistics against the number of LUT nodes $m_i$, fitted by a double exponential function. For comparison, we also show the performance obtained after a homogeneous pseudo-random node distribution following the Sobol's sequence~\cite{Bratley198888}. 

\section{Results}
\label{sec:results}

\begin{figure*}[!ht]
	\centering
	\IG[width=0.45\linewidth,trim=0 1.3cm 26.8cm 0.2cm,clip=true]{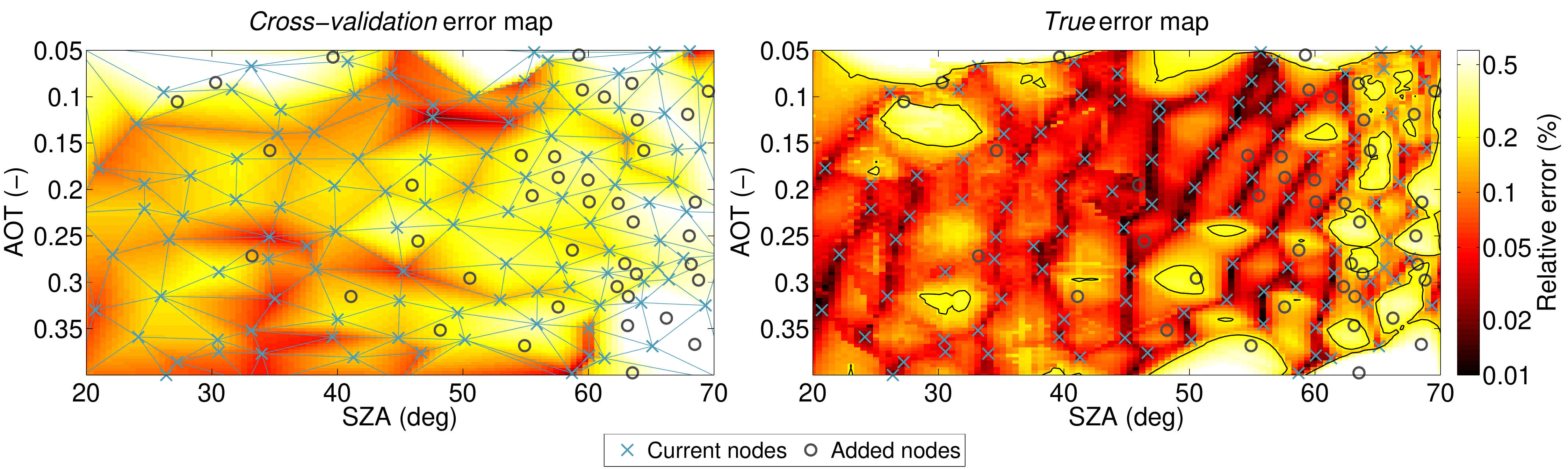}
	\IG[width=0.45\linewidth,trim=0 1.3cm 26.8cm 0.2cm,clip=true]{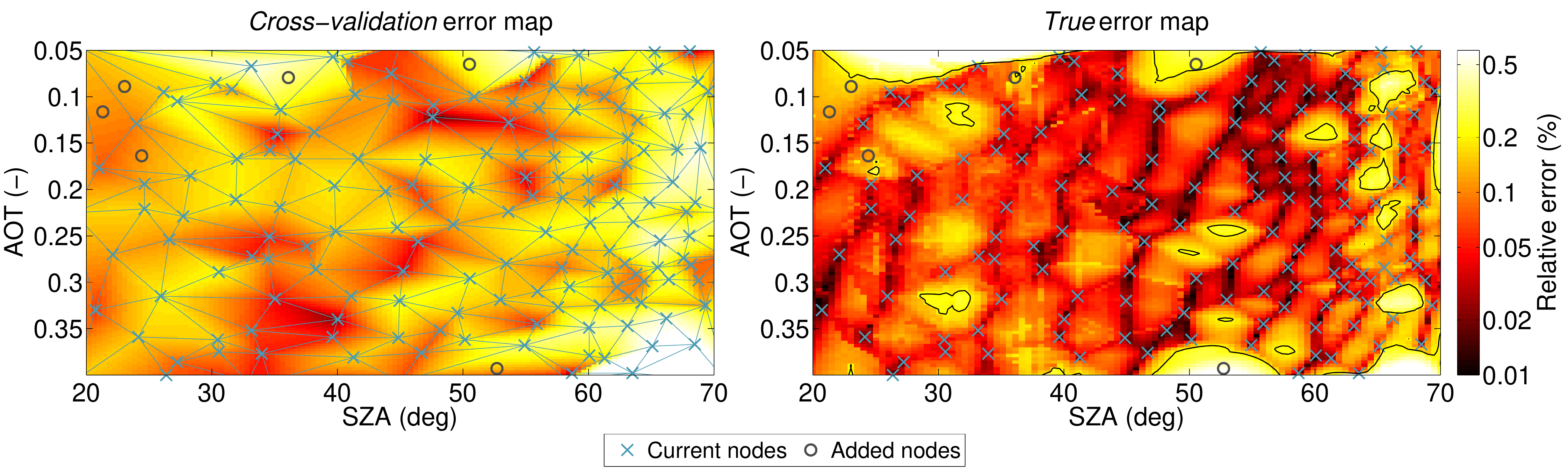}
	\IG[width=0.07\linewidth,trim=46.4cm 1.3cm 0 0.2cm,clip=true]{figs/valCase1_5.pdf}
	\IG[width=\linewidth,trim=0 0.2cm 0cm 13.7cm,clip=true]{figs/valCase1_5.pdf}
	\caption{{\it Cross-validation} error maps for the Case \#1 test at iterations $i=5$ (left) and $i=6$ (right) illustrating respectively the functioning of the {\it geometry} and {\it density} terms of the acquisition function. The light blue lines indicate the underlying Delaunay triangulation.}
	\label{fig:Case1maps}
\end{figure*}

\begin{figure*}[!ht]
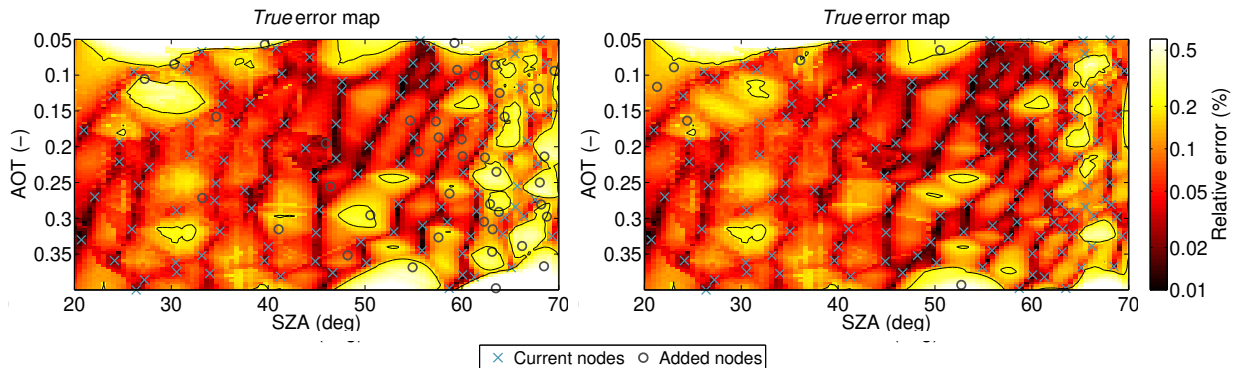

	\centering
	\IG[width=0.45\linewidth,trim=23.2cm 1.3cm 3.7cm 0.2cm,clip=true]{figs/valCase1_5.pdf}
	\IG[width=0.45\linewidth,trim=23.2cm 1.3cm 3.7cm 0.2cm,clip=true]{figs/valCase1_6.pdf}
	\IG[width=0.07\linewidth,trim=46.4cm 1.3cm 0 0.2cm,clip=true]{figs/valCase1_5.pdf}
	\IG[width=\linewidth,trim=0 0.2cm 0cm 13.7cm,clip=true]{figs/valCase1_5.pdf}
	\caption{{\it True} error maps for the Case \#1 test at iterations $i=5$ (left) and $i=6$ (right) illustrating respectively the functioning of the {\it geometry} and {\it density} terms of the acquisition function. The black contour lines indicate the error threshold of $\varepsilon_t$=0.2\%.}
	\label{fig:Case1maps2}
\end{figure*}

First we visualize the functioning of GALGA through the 2$D$ error maps from the test case \#1 (see Figures \ref{fig:Case1maps} and \ref{fig:Case1maps2}). For the actuation of the {\it geometry} term (iteration $i=5$), the new nodes are added in areas where the interpolation error is estimated to be higher than the $\varepsilon_t$=0.2\% error threshold (see {\it cross-validation} error map at Figure \ref{fig:Case1maps}-left). Most of these nodes are located in areas of low TOA radiance (i.e., at SZA$>$60 deg), thus where higher relative interpolation errors are expected. The addition of these new nodes reduce the areas with errors above the threshold as observed in the change of the {\it true} error map between iterations $i=5$ and $i=6$ (see Figure \ref{fig:Case1maps2}). This indicates that method is functioning correctly under the {\it geometry} term. 
Since GALGA approximates the interpolation error based on the ``leave-one-out'' cross-validation technique, we can also observe that the {\it cross-validation} error map has systematically higher error values than the {\it true} error map. Consequently, GALGA leads to an oversampling or undersampling of some areas of the input variable space. On the one hand, some areas have a {\it true} interpolation error at $i=5$ that is already below the error threshold (e.g., SZA$\approx$45 deg and AOT=0.2-0.25). However, GALGA adds new nodes, leading to a local oversampling of the input space (see Figures \ref{fig:Case1maps} and \ref{fig:Case1maps2} left). On the other hand, undersampled areas (e.g., at SZA$\approx$52 deg and AOT$\approx$0.4) still remain with a high interpolation error (see Figure \ref{fig:Case1maps2}-right). The {\it density} term of the acquisition function intends to reduce the amount of undersampled areas. Indeed, at iteration $i=6$ the new added nodes are located in barycenter of the simplices with largest (undersampled) areas. 
Additionally for this particular case, we can observe a pattern of low interpolation errors connecting nodes with similar SZA but different AOT (see dark red vertical pattern in the {\it true} error map). This indicates that linear interpolation derives larger errors when interpolating between SZA values than between AOT values.

 
\begin{figure}[h!]
	\centering
	\IG[width=\linewidth,trim=0.2cm 0.4cm 0.2cm 0.3cm,clip=true]{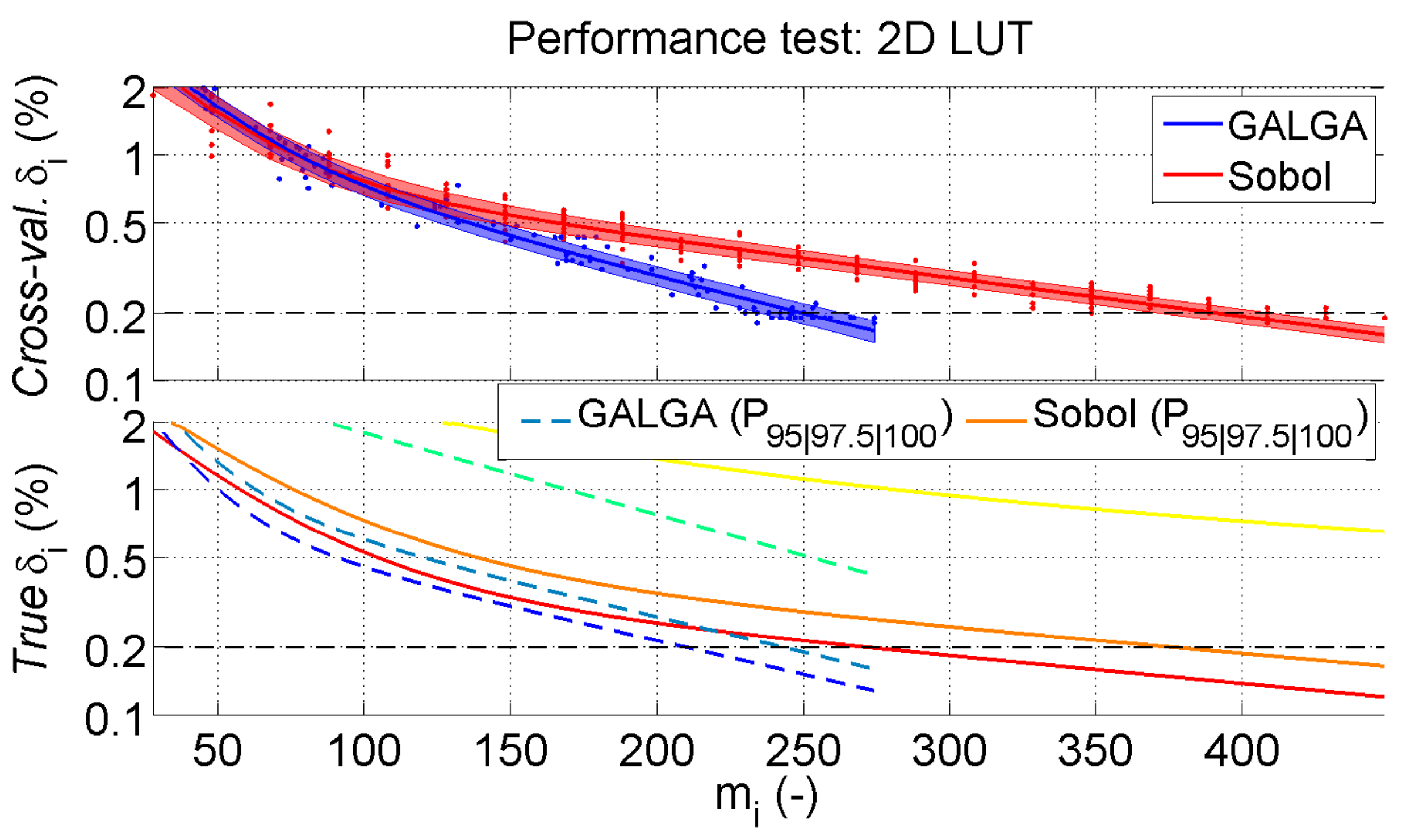}
	\caption{Estimated (top) and true (bottom) performance of GALGA (blueish colors) and Sobol distribution (reddish colors) in a 2$D$ LUT (Case \#1). For the top panel, mean (solid lines) and standard deviation (shaded areas) are obtained after averaging $N$=10 independent runs. In the bottom panel, three percentile values (95\%, 97.5\% and 100\%) of the interpolation error histograms are represented with a different color shade (darker to lighter). The error threshold, $\varepsilon_t$, is indicated with the horizontal dashed black line at 0.2\%.}
	\label{fig:2Da}
\end{figure}

We continue by assessing the performance of the proposed method against a Sobol pseudo-random homogeneous distribution of LUT nodes. The analysis is done for LUTs of increasing input dimensions: 2$D$, 4$D$ and 6$D$. When evaluating the algorithm performance for the Case \#1 (2$D$ LUT; see Figure \ref{fig:2Da}), we can observe that the gradient-based automatic LUT generator method outperforms the accuracy obtained with the Sobol pseudo-random distribution after $m_i\approx$150 nodes. In terms of the estimated performance (see Figure \ref{fig:2Da}-top), our method needs approximately $m_i$=250 nodes to achieve the required error, reducing the LUT size with respect to a Sobol distribution down to 67\% ($m_i$=375 nodes). Our method not only reduces the LUT size but also gets lower interpolation error (see Figure \ref{fig:2Da}-bottom) after nearly $m_i$=175 nodes. This happens for all the percentiles between 95\% and 100\%. In fact, with the complete LUT of $m_i\approx$250 nodes, our method reaches an interpolation error below the $\varepsilon_t$=0.2\% error threshold in nearly 97.5\% of the input variable space (maximum error $\sim$0.5\%). Instead, a LUT constructed with a Sobol pseudo-random distribution reaches the $\varepsilon_t$=0.2\% error threshold in only $\sim$95\% of the input space (maximum error $\sim$2\%).

\begin{figure}[h!]
	\centering
	\IG[width=\linewidth,trim=0.2cm 0.4cm 0cm 0.3cm,clip=true]{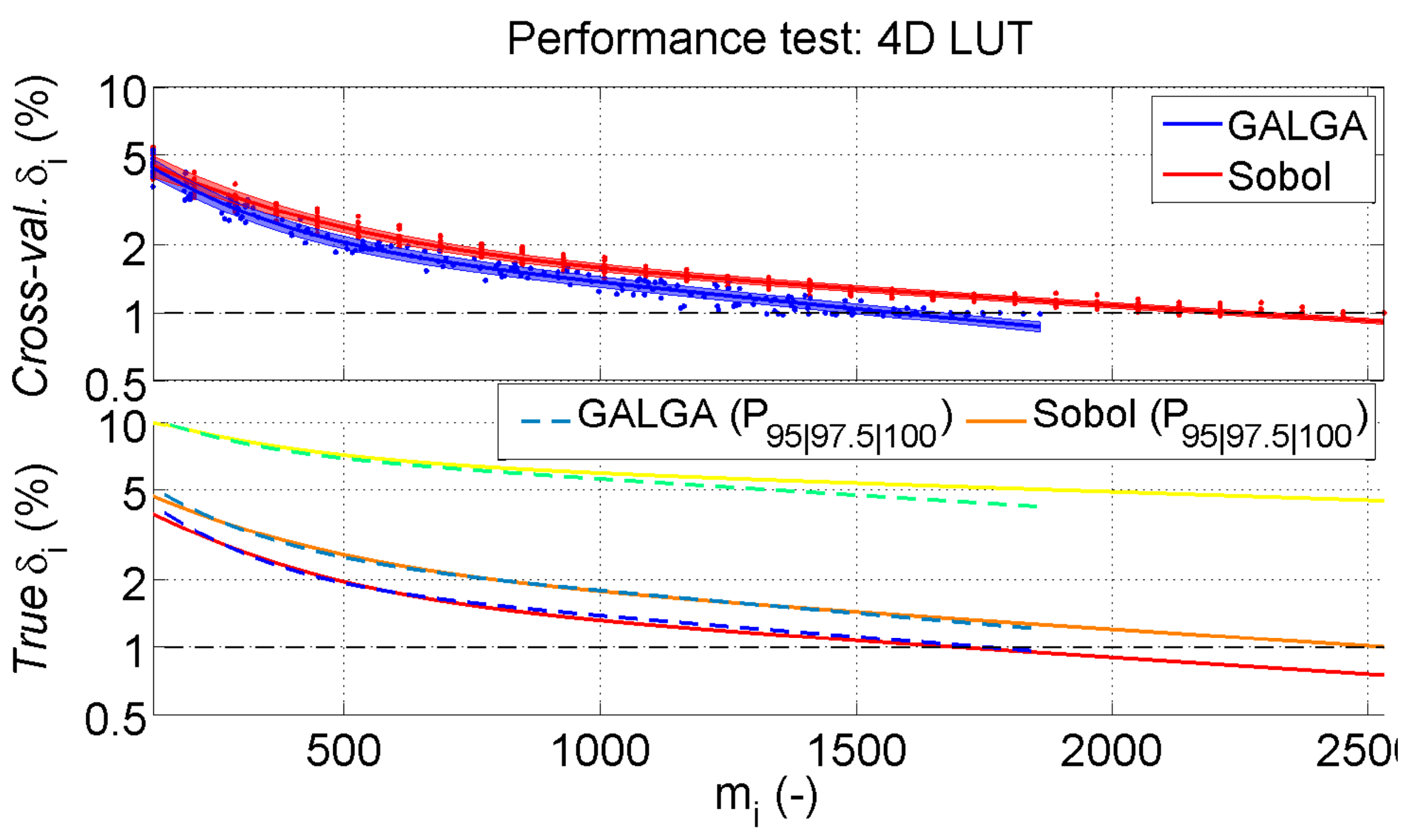}
	\caption{Estimated (top) and true (bottom) performance of GALGA (blueish colors) and Sobol distribution (reddish colors) in a 4$D$ LUT (Case \#2). For the top panel, mean (solid lines) and standard deviation (shaded areas) are obtained after averaging $N$=10 independent runs. In the bottom panel, three percentile values (95\%, 97.5\% and 100\%) of the interpolation error histograms are represented with a different color shade (darker to lighter). The error threshold, $\varepsilon_t$, is indicated with the horizontal dashed black line at 1\%.}
	\label{fig:4D}
\end{figure}

When evaluating the algorithm performance for the Case \#2 (4$D$ LUT; see Figure \ref{fig:4D}), we observed that, according to the cross-validation error, the proposed method is still performing better than a pseudo-random homogeneous distribution from already 500 nodes. Through the evaluation of the cross-validation error, nearly $m_i$=1700 nodes are needed with the distribution proposed in our method to reach an interpolation error of 1\% in 95\% of the cases, i.e., 74\% lower with respect to a Sobol distribution (nearly $m_i$=2300 nodes). However, the evaluation of histogram of the true error (bottom plot) shows that, for most of the points in the input space (95\% and 97.5\% percentiles) both distribution methods obtain the same interpolation error. Only when analyzing the errors in the higher part of the histogram (percentiles $>$98\%) we observe that the proposed method achieve superior accuracies than with a homogeneous Sobol distribution. 

\begin{figure}[h!]
	\centering
	\IG[width=\linewidth,trim=0.2cm 0.4cm 0cm 0.3cm,clip=true]{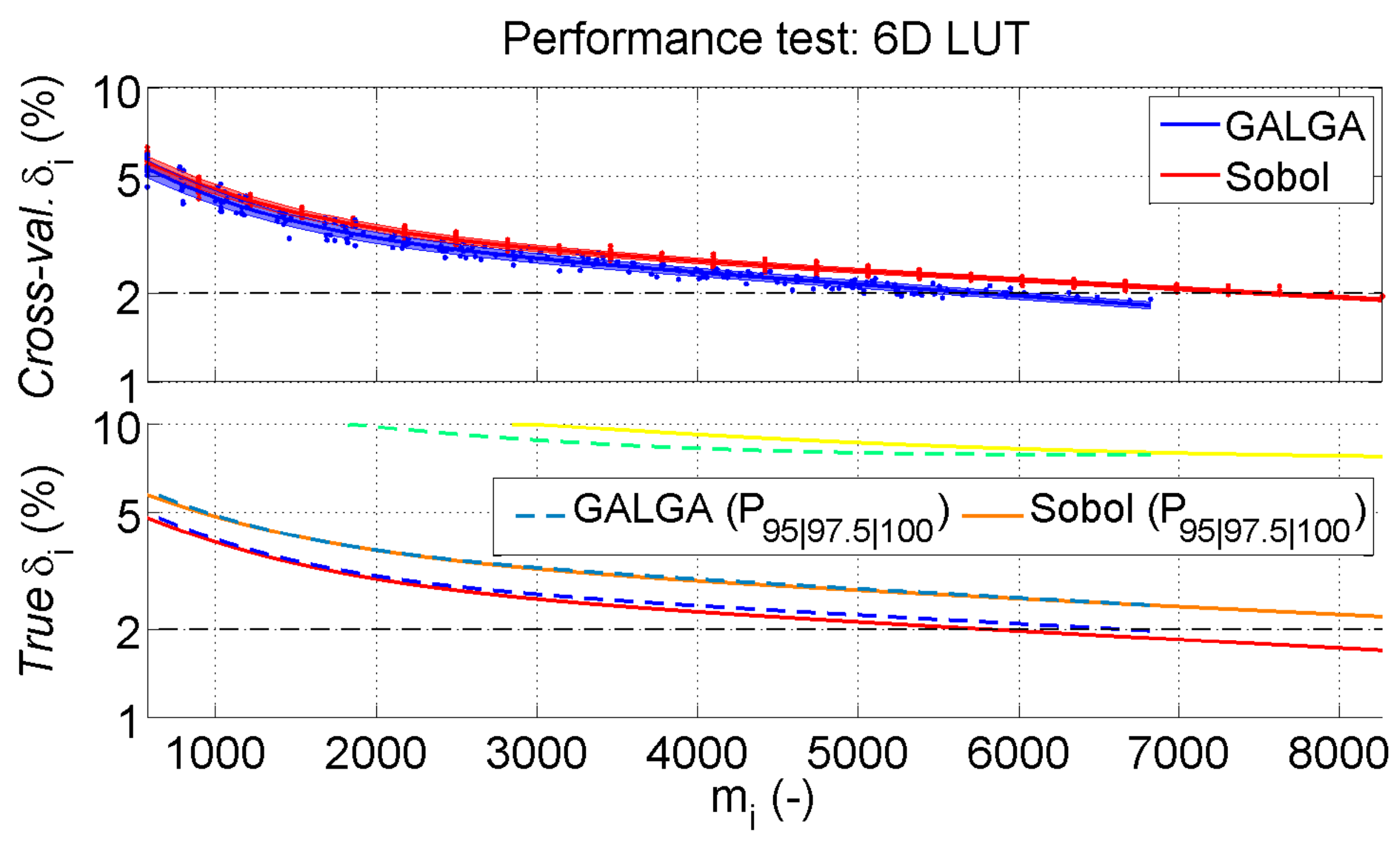}
	\caption{Estimated (top) and true (bottom) performance of GALGA (blueish colors) and Sobol distribution (reddish colors) in a 6$D$ LUT (Case \#3). For the top panel, mean (solid lines) and standard deviation (shaded areas) are obtained after averaging $N$=10 independent runs. In the bottom panel, three percentile values (95\%, 97.5\% and 100\%) of the interpolation error histograms are represented with a different color shade (darker to lighter). The error threshold, $\varepsilon_t$, is indicated with the horizontal dashed black line at 2\%.}
	\label{fig:6D}
\end{figure}

As we increase the dimensionality of the input variable space, we observe the same trend in the algorithm performance. For the Case \#3 (6$D$ LUT; see Figure \ref{fig:6D}), the evaluation of the cross-validation error indicates that the performance of our method is better than the Sobol distribution, which is clearly seen after nearly $m_i$=3000 LUT nodes. Our method achieves an interpolation error of 2\% for $m_i$=5500 nodes, which is 77\% less that the nodes needed with a pseudo-random homogeneous distribution ($m_i$=7200 nodes). However, the evaluation of the true interpolation error when compared with the reference LUT indicates that both node distribution methods achieve nearly the same accuracy. Only for the maximum interpolation errors (percentiles 100\%), our method obtains slightly lower interpolation errors than with the Sobol node distribution.

\section{Conclusions \& Outlook}
\label{sec:conclusions} 
In this work, we have proposed GALGA, a new method to optimize the node distribution of multi-dimensional LUTs. Particularly, the proposed algorithm is applied here to the construction of MODTRAN LUTs of {\it atmospheric transfer functions} in order to reduce (1) errors in the interpolation of these atmospheric transfer functions, and (2) computation time to build these LUTs. 
The proposed method is based on the exploitation of the gradient/Jacobian information of the underlying function (TOA radiance in our case) and the concept of an acquisition function, divided into its {\it geometry} and {\it density} terms. Through the experimental set-up, we have verified that the algorithm functions as expected, observing that the use of the acquisition function identifies areas in the input variable space with high interpolation errors. 
 
Thus, the proposed method reduces the number of nodes needed to construct a LUT by nearly 75\% of the nodes needed using a pseudo-random homogeneous distribution. The performance of GALGA was also evaluated by calculating the real interpolation error in LUTs of 2$D$, 4$D$ and 6$D$. The LUTs constructed with the proposed method achieve an interpolation error that is, in the worst case, equivalent to the interpolation error obtained with a LUT of homogeneously distributed nodes. The largest interpolation relative errors are also reduced by 0.5\% with LUTs designed GALGA when compared against those obtained with a Sobol distribution.
However, there is an {apparent low gain in true accuracy observed in the 4$D$ and 6$D$ cases and that might be explained by two factors.}
On the one hand, the algorithm takes several iterations to have a density of the LUT input variable space that is enough to identify areas with higher sensitivity to interpolation errors. For the selected interpolation error threshold ($\varepsilon_t$), these specific areas still might represent a small portion of the input variable space.
On the other hand, the number of nodes in the {\it ground truth} LUTs might be insufficient to have a representative discrete sampling of the underlying TOA radiance in the input variable space]. This low sampling causes that just a few {\it ground truth} LUT nodes are distributed in areas where GALGA gives a gain in accuracy, falsely increasing the accuracy obtained with the Sobol distribution.
For these two factors, both LUT node distribution methods (i.e., GALGA and Sobol) obtain similar histogram of the interpolation error. 

GALGA has been implemented in the {\it Atmospheric LUT Generator (ALG)} v1.2 software~\cite{Vicent2017}. ALG allows generating LUTs based on a suite of atmospheric RTMs, facilitating consistent and intuitive user interaction, thereby streamlining model setup, running and storing RTM data for any spectral configuration in the optical domain. In combination with ALG, GALGA facilitates users generating optimized atmospheric LUTs, reducing computation time in the execution of atmospheric RTMs and improving the accuracy of LUT interpolation. The proposed algorithm can eventually be implemented for the generation of LUTs in a wider range of remote sensing applications, including vegetation and water RTMs~\cite{Jacquemoud2009,hydrolight}. Compact and informative LUTs give rise to interesting possibilities such as optimization of biophysical parameters retrieval algorithms~\cite{Verrelst14}, atmospheric correction~\cite{Guanter2009} and RTM emulation~\cite{Rivera2015,Verrelst2016}.

Future research will focus on the use of statistical methods to improve the reconstruction of the underlying interpolation error in the TOA radiance, which have been demonstrated to be suitable for atmospheric RTM~\cite{Vicent2017b}. Therefore, our previous work in the {\em AGAPE} algorithm~\cite{luca2017a} will be expanded for the multi-output (i.e., spectral) RTM output data. Altogether, we are aiming at further optimizing the distribution of LUT nodes and reducing the errors in LUT interpolation.

\end{document}